# Quantum-mechanical effect in atomically thin $MoS_2$ FET

**Nan Fang[1,2] and Kosuke Nagashio[1]**

[1]Department of Materials Engineering, The University of Tokyo, Tokyo 113-8656, Japan
[2]Present address: RIKEN, Hirosawa, Wako, Saitama 351-0198, Japan
E-mail: nagashio@material.t.u-tokyo.ac.jp

**Abstract:** Two-dimensional (2D) layered materials-based field-effect transistors (FETs) are promising for ultimate scaled electron device applications because of the improved electrostatics to atomically thin body thickness. However, compared with the typical thickness of ~5-nm for Si-on-insulator (SOI), the advantage of the ultimate thickness limit of monolayer for the device performance has not been fully proved yet, especially for the on-state at the accumulation region. Here, we present much stronger quantum-mechanical effect at the accumulation region based on the *C*-*V* analysis for top-gate $MoS_2$ FETs. The self-consistent calculation elucidated that the electrons are confined in the monolayer thickness, unlike in the triangle potential formed by the electric field for SOI, the gate-channel capacitance is ideally maximized to the gate insulator capacitance since the capacitive contribution of the channel can be neglected due to the negligible channel thickness. This quantum-mechanical effect agreed well with the experimental results. Therefore, monolayer 2D channels are suggested to be used to enhance the on-current as well as the gate modulation ability.

## 1. Introduction

The metal-oxide-semiconductor field-effect transistor (MOSFET) scaling requires the transition from planar to FinFET to overcome the short channel effect, which prolongs the life of the silicon complimentary MOS [1,2]. The electrostatics have been improved, while 3-dimensional (3D) shape has drastically increased the parasitic capacitance ($C_{para}$) to the level beyond the intrinsic gate capacitance, which results in the loss of gate controllability [3]. Therefore, it has been suggested that an attractive charge-based device scaling path should be back to 2D from 3D to reduce the parasitism [4].

So far, for the conventional 2D channel, i.e., silicon-on-insulator (SOI) MOSFETs, the electrostatic field-effect control of carriers has already been investigated by deepening the understanding of the bulk Si MOSFET, which is characterized by capacitance-voltage (*C*-*V*) measurements [5]. The bulk Si MOSFET usually operates at the inversion mode, and the carriers in the inversion layer are confined by the electrical-field-induced band bending, as shown in **Figure 1**. In the classical simplest model, the carriers are assumed to stay just at the $SiO_2$/Si interface. The gate-channel capacitance ($C_{gc}$) is equal to the oxide capacitance ($C_{ox}$). However, the electrons are governed by quantum mechanics and described by wave functions. Therefore, the carriers in the electrical-field-induced band bending are quantized into a 2D subband structure, where $E_1$ and $E_2$ correspond to the two lowest energy subbands. The distance between the centroid for the square modulus of the wave function and the interface is shown by $z_{inv}$. Therefore, the additional contribution resulted from this distance in the semiconductor should be considered and called the quantum mechanical effect. This results in $1/C_{gc} = 1/C_{ox} + 1/C_{inv}$, where $C_{inv}$ is the inversion layer capacitance. For thin-channel SOI MOSFETs (*e.g.*, 5-nm-thick silicon), the quantized carriers are further confined by the channel thickness, which results in the sharper carrier distribution and slightly enhances $C_{gc}$ [5]. However, the *C-V* analysis at the atomic thickness limit, which is the case for the 2D layered channels, has not been experimentally demonstrated for the SOI MOSFET, because there is a drastic mobility degradation when the Si thickness decreases from 5 nm to 1 nm [6], and the fabrication of SOI with a uniform thickness of 1 nm is extremely difficult [6,7].

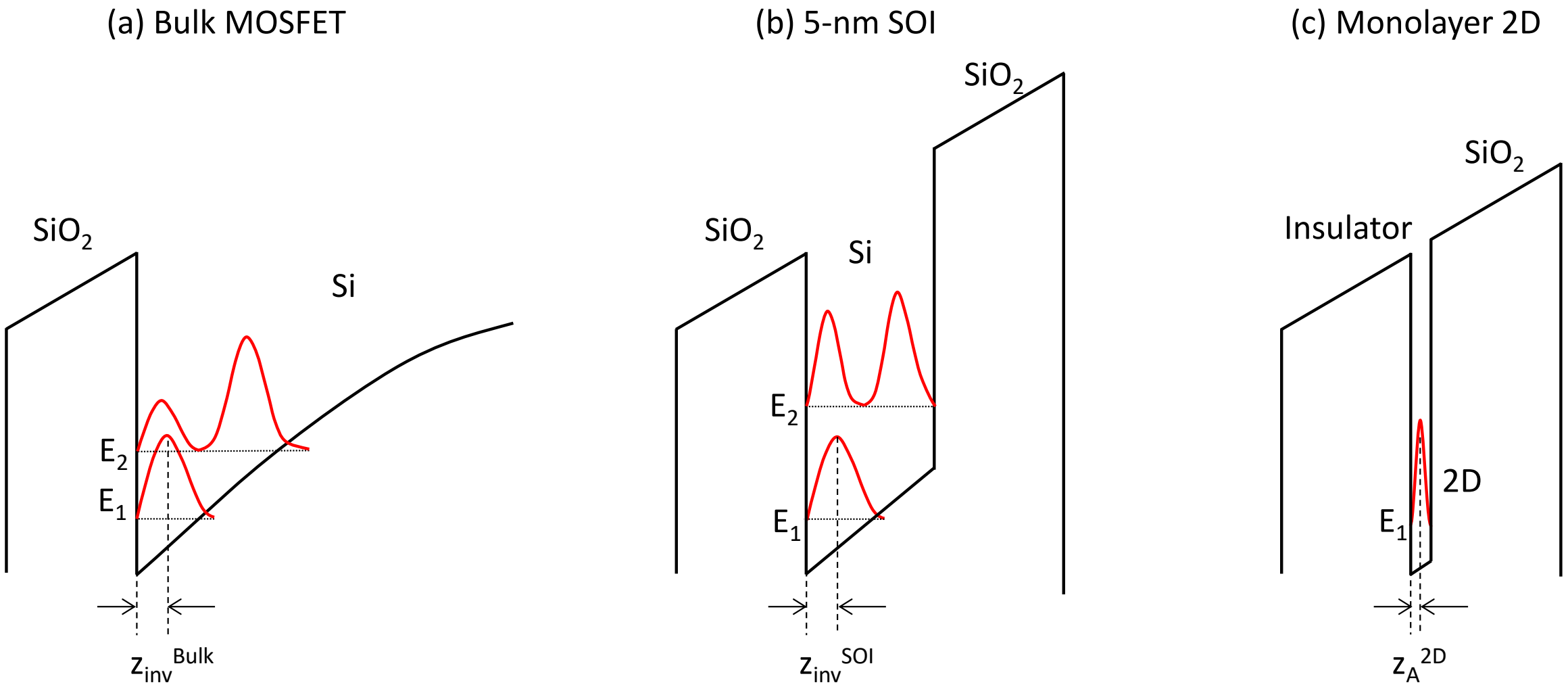

**Figure 1** Schematic illustration of the subband structures for bulk Si-MOSFET, 5-nm SOI, and monolayer 2D. Red lines indicate square modulus of the wave functions for subbands. Si-MOSFET and SOI operate at the inversion mode, while 2D FET operates at the accumulation mode. Therefore, the distance from the interface to the center of the square modulus of the wave function is expressed as $z_{inv}$ and $z_A$, respectively.

For atomically thin 2D layered channels, *C-V* measurements were reported on $MoS_2$ FETs to analyze the interface state densities [8-15], electrical reliability on the gate oxide [16-18], quantum capacitance ($C_Q$) [8], metal-insulator transition [19] and thickness-dependent depletion behavior [20]. It has been recognized that most of 2D-layered-channel-based FETs operate at the accumulation mode [20]. Therefore, the strong quantum-mechanical effect is expected at the accumulation region, which cannot be achieved in the SOI MOSFET.

In this work, the accumulation region in *C-V* for top-gate $MoS_2$ FETs is systematically analyzed to elucidate the quantum-mechanical effect. In addition, the self-consistent calculation is performed by solving Poisson equation and Schrödinger equation. Finally, based on the comparison of the experimental and theoretical results, the quantum-mechanical effects on $C_{gc}$ are discussed from the viewpoint of the 2D device thickness scaling.

## 2. Results and discussion

In this paper, $MoS_2$ films were mechanically exfoliated onto the quartz substrate from natural bulk $MoS_2$ flakes. Raman spectroscopy and atomic force microscopy (AFM) were employed to determine the layer number. Ni/Au was deposited as the source/drain electrodes. Then, Y metal with a thickness of 1 nm was deposited via thermal evaporation of the Y metal from a PBN crucible in an Ar atmosphere with a partial pressure of $10^{-1}$ Pa, followed by oxidization in the laboratory atmosphere [21,22]. $Y_2O_3$ acts as the buffer layer for the nucleation sites on $MoS_2$ because $MoS_2$ surface is chemically inert. The $Al_2O_3$ oxide layer with a thickness of 10 nm was deposited via atomic layer deposition (ALD), followed by the Al top-gate electrode formation. *C-V* measurements were conducted using 4980A LCR meters in a vacuum prober at the room temperature.

**Figure 2** shows a schematic drawing and an optical image of the top-gate $MoS_2$ FET. The $MoS_2$ flakes with the large area (>30 $\mu m^2$) were selected for device fabrication and characterization because the measured capacitance should be larger than the stray capacitance (~10 fF) of the measurement system. Moreover, it should be emphasized that the quartz substrate was used to remove $C_{para}$ completely [20]. $C_{gc}$ was measured in the top-gate $MoS_2$-FETs, where the top-gate electrode was connected to the high terminal, while both source and drain were connected to the low terminal. The full equivalent circuit of the $MoS_2$ accumulation region is shown in **Figure 2b**. $C_A$ is the accumulation capacitance. $C_{it}$ and $R_{it}$ are the capacitance and resistance of the interface states, respectively, which account for the carrier capture and emission processes. $R_{ch}$ is the channel resistance, which accounts for the carrier charging process to the channel through the source/drain. At the depletion region, $C_{it}$ is important in the *C-V* characteristics because $C_{it}$ is much larger than the depletion capacitance, while it can be neglected at the strong accumulation region in this study. $R_{ch}$ becomes dominant at the depletion region but can be neglected at the strong accumulation region due to the high carrier density. More detailed information on the equivalent circuit can be found elsewhere [8,20]. Therefore, at the accumulation region, the simplified equivalent circuit with $C_A$ and $C_{ox}$ can be considered, as shown in **Figure 2c**.

**Figure 3a** shows the experimental $C_{gc}$-$V_{TG}$ curves for the $MoS_2$ channel thicknesses ($t_{ch}$) of 1 layer (1L), 3L, and 10 nm at the frequency of 1 MHz. The 1L $MoS_2$ shows (i) the highest saturation capacitance and (ii) a sharp rise at the accumulation region, while 10-nm $MoS_2$ shows the lowest saturation capacitance and a slow rise to the accumulation region. Since all devices have identical $C_{ox}$ in principle due to the identical top-gate process, the thickness dependence of $C_{gc}$ can arise from the contribution of $C_A$. Here, in the case of the SOI MOSFET, the sharp rise of the capacitance was observed with the decrease in the thin body thickness [5]. However, the increase in the capacitance at the saturation region was not

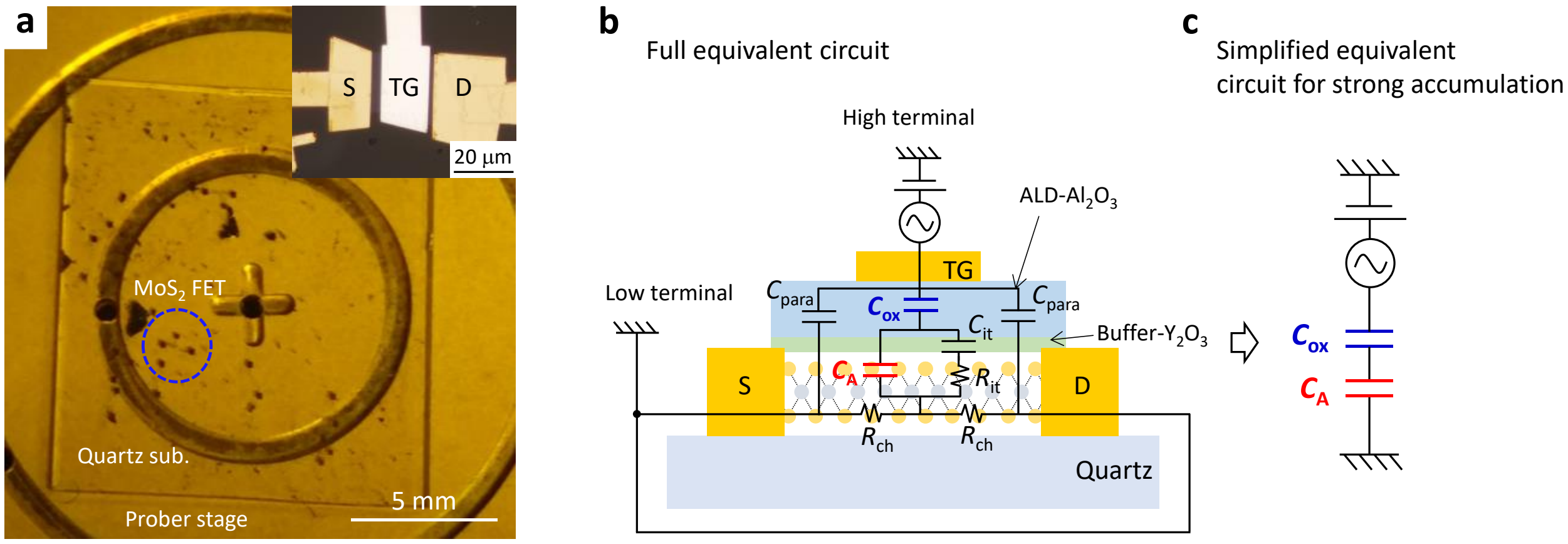

**Figure 2** (a) Whole image of the devices on a quartz substrate. The inset is the magnified image showing a top-gate $MoS_2$ FET. (b) Schematics of the $MoS_2$ FET structure and full equivalent circuit of the capacitance measurements. (c) Simplified equivalent circuit at the strong accumulation region.

observed. The drastic increase in capacitance at the saturation region in the present $MoS_2$ FET can be attributed to the giant quantum-mechanical effect at the monolayer limit, which is quite difficult to experimentally achieve by the channel thickness scaling of the SOI MOSFET.

The self-consistent calculation was performed by solving Poisson equation and Schrödinger equation to clarify the quantum-mechanical effect in the $MoS_2$ FET. **Figure 3c** shows the example of the energy band diagram of the top $Al_2O_3$ (10 nm)/$MoS_2$ (10 nm)/back $Al_2O_3$ (10 nm) gate stack used for the calculation. The gate voltage is always applied to the left $Al_2O_3$ oxide, which is used as the top-gate oxide. The right $Al_2O_3$ oxide is just considered as the substrate, and the effect of the thickness and oxide type on the calculated results can be ignored here. Schrödinger equation was solved in the effective mass approximation using the software by incorporating material properties specific to the device [23]. The parameters of $MoS_2$ used in the calculation are shown in **Table 1**. The thickness of $MoS_2$ is varied from 1L to 10 nm at different gate biases. For simplicity, the continuous band is used instead of the discrete bands with the van der Waals energy gap. The layer number dependence of the dielectric constant ($\varepsilon_{ch}$) for $MoS_2$ is adapted from the studies of first-principle calculation [24, 25]. This simplification is reasonable, since it provides a sufficient explanation of the experimental results. The calculated $C_{gc}$-$V_{TG}$ curves are shown in **Figure 3b**. The 1L $MoS_2$ shows a sharp rise and the highest saturation capacitance at the accumulation region, which is consistent with the experimental results.

**Table 1** Physical parameters used in the simulation.

| | Dielectric constant | Effective mass | $N_D$ ($cm^{-3}$) |
|---|---|---|---|
| monolayer | 3.0 | 0.73 | $3\times10^{17}$ |
| bilayer | 4.2 | 0.73 | $3\times10^{17}$ |
| bulk | 6.3 | 0.73 | $3\times10^{17}$ |

Here, the origin for the thickness dependence of $C_{gc}$ is discussed. **Figures 4a-c** show the calculated three lowest subband energies ($E_1$, $E_2$, $E_3$), square modulus of wave function for each subband, and channel electron density ($n$) distribution for 10-nm $MoS_2$ at the

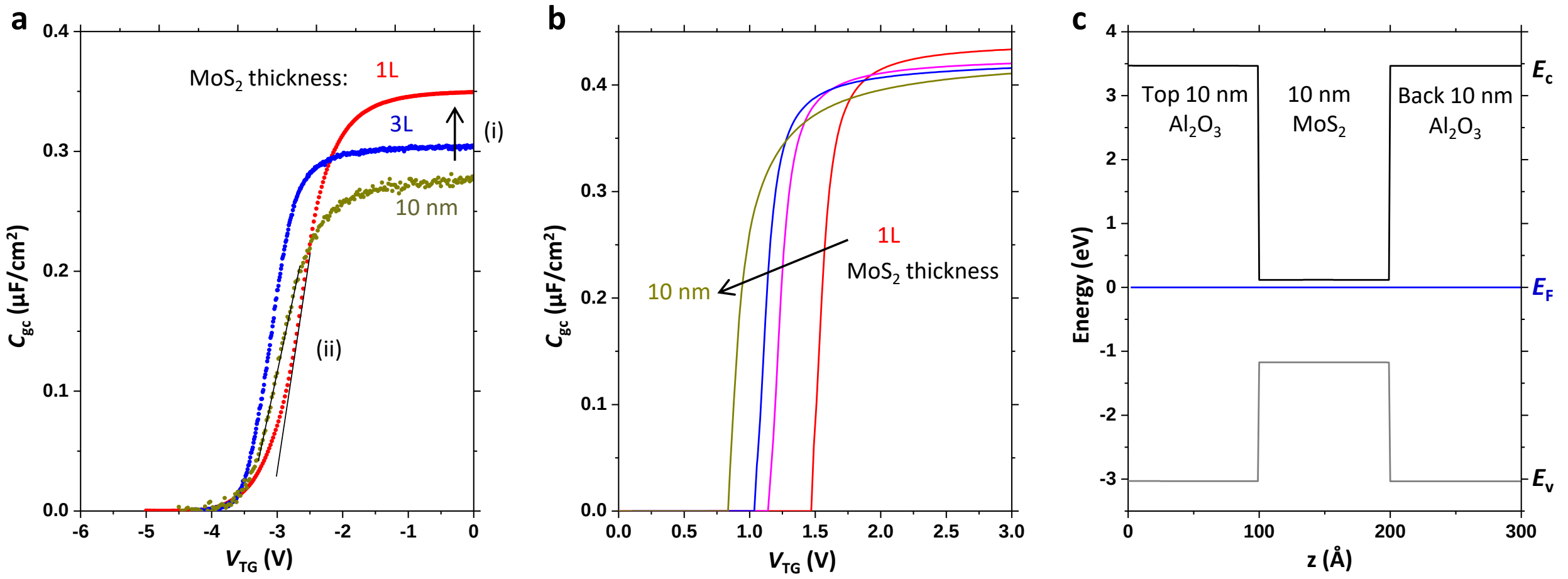

**Figure 3** (a) Experimental $C_{gc}$-$V_{TG}$ characteristics at 1 MHz of $MoS_2$ FETs with a thickness of 1L, 3L, and 10 nm. (b) Simulated $C_{gc}$-$V_{TG}$ characteristics of the $MoS_2$ FETs with a thickness of 1L, 2L, 3L, and 10 nm, respectively. The solid lines are the results from the self-consistent calculation by solving Poisson equation and Schrödinger equation [23]. (c) Typical example of the energy band diagram of the top $Al_2O_3$ (10 nm)/$MoS_2$ (10 nm)/back $Al_2O_3$ (10 nm) gate stack. $E_C$, $E_V$, and $E_F$ are conduction band, valance band edge energies, and Fermi energy, respectively.

weak depletion. The energy difference between the subbands ($E_2$ - $E_1$ = 0.017 eV) is smaller than the $k_BT$ ($k_B$ and $T$ are defined as the Boltzmann constant and the temperature), which makes all of these subbands occupied by the electrons. As a result, $n$ is distributed throughout the entire channel thickness with a peak near the back surface. Meanwhile, when the positive gate bias is applied, the strong accumulation is achieved as shown in **Figures 4d-f**. The energy difference between $E_1$ and $E_2$ increases due to the confinement in the top surface triangle potential formed by the electrical field. As a result, most of electrons locate in the lowest energy subband $E_1$, and the wave function of $E_1$ is confined in the top surface triangle potential. That is, 2D electron gas (2DEG) is formed near the top surface of $MoS_2$ with the distance of ~1.1 nm. Therefore, when the gate bias is modulated from the depletion to the accumulation, the centroid position for the square modulus of the wave function is gradually changed from the bottom to the top of the $MoS_2$ body, which results in a slow rise of $C_{gc}$ toward the accumulation region by the gate voltage.

When $t_{ch}$ is reduced, in addition to the electrical field confinement, the thickness confinement will dominate the quantum-mechanical effect. It is similar to the consideration of a quantum well [26], and the

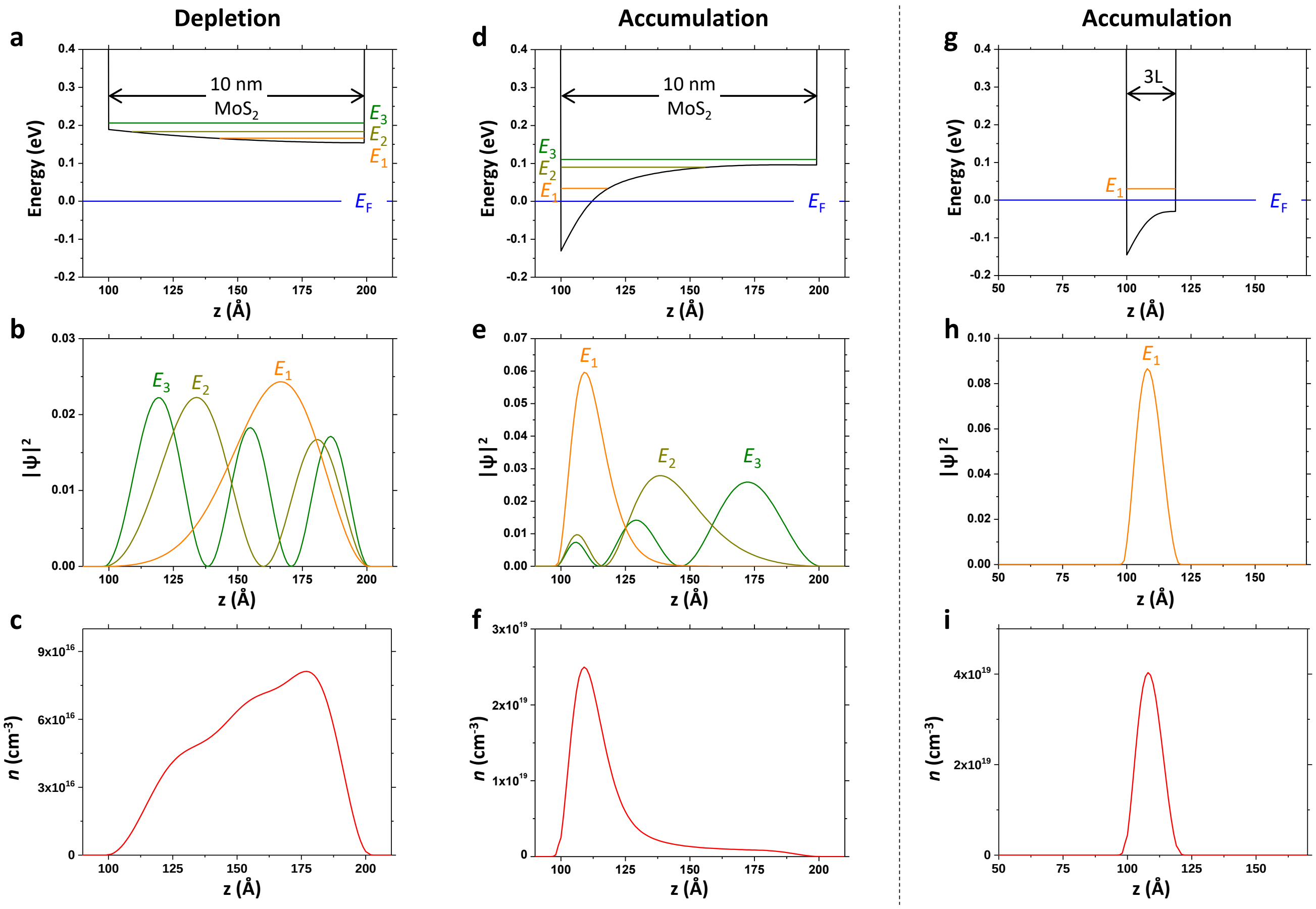

**Figure 4** (a) Energy band diagram of the 10-nm $Al_2O_3$/10-nm $MoS_2$/10-nm $Al_2O_3$ structure at the weak depletion region. The gate voltage is applied on the left $Al_2O_3$ oxide. (b) Square modulus of the wave function of each subband. (c) Channel electron density distribution. The strong accumulation case for the 10-nm $Al_2O_3$/10-nm $MoS_2$/10-nm $Al_2O_3$ structure is shown in (d-f). The strong accumulation case of the 10-nm $Al_2O_3$/3L $MoS_2$/10-nm $Al_2O_3$ structure is shown in (g-i).

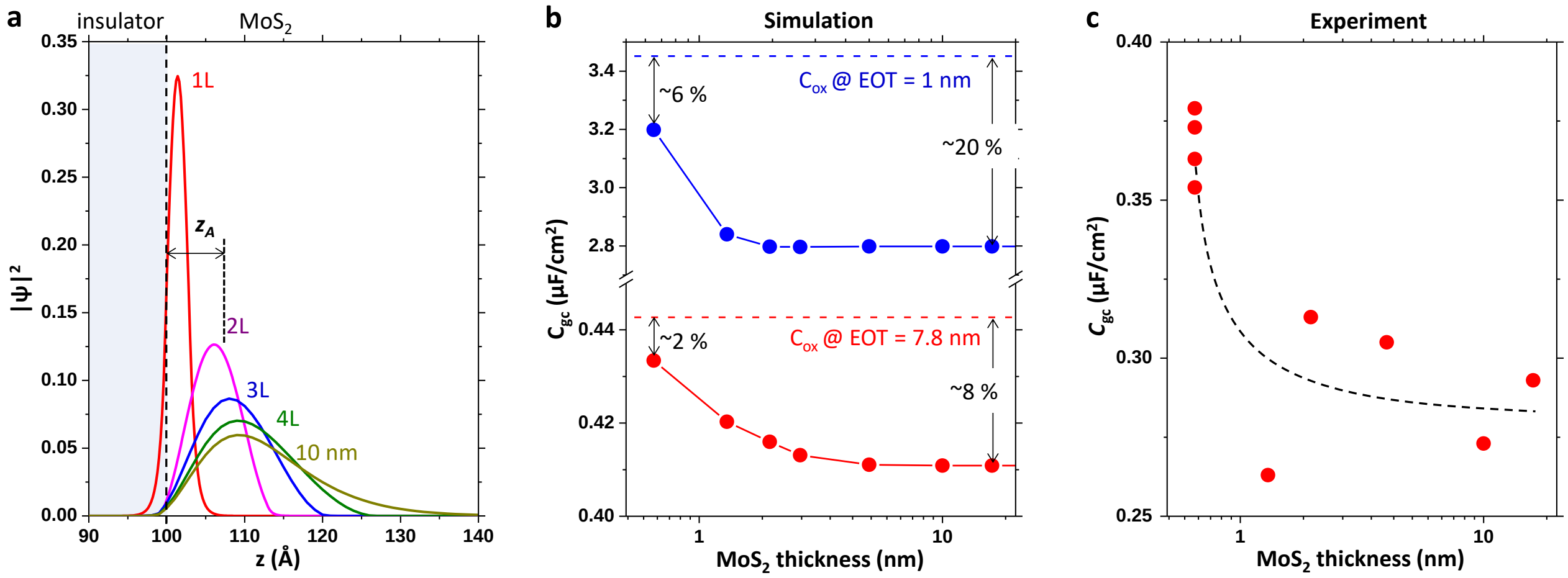

**Figure 5** (a) Square modulus of the wave function of the lowest subband ($E_1$) with the $MoS_2$ thickness of ~ 1L, 2L, 3L, 4L, and 10 nm at the strong accumulation region. (b) Simulated $C_{gc}$ for EOT = 1 nm and 7.8 nm and (c) experimental $C_{gc}$ at the strong accumulation region as a function of the $MoS_2$ thickness.

thickness-confined subband energy eigenvalues can be estimated as $E_n = \hbar\pi^2 n^2 / 2m_{ch}t_{ch}^2$, where $m_{ch}$ is the effective mass of $MoS_2$. By decreasing $t_{ch}$, the energy difference between $E_1$ and $E_2$ will be much larger than $3k_BT$, and only subband $E_1$ can be considered to be occupied. As a result, the intrinsic 2DEG is formed in the ultrathin $MoS_2$ channel. This is the scenario for 3L to 1L $MoS_2$, as shown in **Figures 4g-i**. Subband $E_1$ is now confined in the $MoS_2$ body. The electrical-field-induced band bending does not affect the electron distribution because $E_1$ is already higher than the triangle potential. The energy difference between $E_1$ and $E_2$ is quite large, so $E_2$ is not visible in **Figure 4g**. Therefore, when the gate bias is modulated from the depletion to the accumulation, the centroid position for the square modulus of the wave function is almost fixed and results in the effective modulation of $n$. This scenario accounts for the sharper rise of the *C*-*V* curves for 3L to 1L $MoS_2$. Moreover, in **Figure 3b**, the positive shift of $C_{gc}$-$V_{TG}$ curves is observed by decreasing $t_{ch}$. This result is attributed to the increase in the $E_1$ subband energy. However, the experimental $C_{gc}$-$V_{TG}$ curves do not show a clear thickness-dependent shift. This is because the threshold voltage for the 2D layered channel is quite sensitive to the surface doping effect due to the extrinsic adsorbates especially for the ultra-thin channel case [27,28].

The next task is to explain why the saturation capacitance drastically increases at the monolayer limit. In analogy to $C_{inv}$ in SOI MOSFETs, $C_{gc}$ in $MoS_2$ FETs at the accumulation region can be divided as $1/C_{gc} = 1/C_{ox} + 1/C_A = 1/C_{ox} + 1/C_A^{DOS} + 1/C_A^{thickness}$ [29,30]. $C_A^{DOS}$ comes from the finite DOS of $MoS_2$ as $C_A^{DOS} = e^2 g_{2D}[1 + exp(E_G/2k_BT)/2cosh(E_F/k_BT)]$, where $g_{2D} = g_s g_v m_{ch}/2\pi\hbar^2$ is the band-edge *DOS* [8,20,31]. $E_G$ is the bandgap, and $g_s$ and $g_v$ are the spin and valley degeneracy factors, respectively. $C_A^{thickness}$ comes from the distance between the centroid for the square modulus of the wave function and the gate insulator interface ($z_A$) as $C_A^{thickness} = \varepsilon_{ch}/z_A$. At the strong accumulation region, the Fermi energy ($E_F$) enters the conduction band, and $C_A^{DOS}$ corresponds to conduction band-edge DOS. $MoS_2$ has a relatively large effective mass, and therefore $C_A^{DOS}$ is large with the order of ~ 80 μF/cm$^2$ [8]. It is much larger than $C_{ox}$ (typically ~0.3-0.4 μF/cm$^2$) and can be neglected. As a result, it is simplified as $1/C_{gc} = 1/C_{ox} + z_A/\varepsilon_{ch}$, which indicates that $z_A$ determines $C_{gc}$. Therefore, the square modulus of the wave function for subband $E_1$ is shown for different $MoS_2$ thicknesses at the strong accumulation region in **Figure 5a**. In the 10-nm $MoS_2$ case, the centroid of the wave function is confined in the triangle potential induced by band bending instead of its body thickness, and $z_A$ is ~ 1.1 nm. By decreasing the $MoS_2$ thickness, the centroid approaches the interface due to the channel body confinement effect. In the monolayer case, $z_A$ is only ~0.2 nm.

From the above theoretical and experimental results, $MoS_2$ channel thickness seriously affects $z_A$, which further affects $C_A$. The contribution of $C_A$ on $C_{gc}$ is discussed quantitatively from the viewpoint of thickness scaling as below. **Figure 5b** shows the calculated $C_{gc}$ as a function of $t_{ch}$. When the equivalent oxide thickness (EOT) is 7.8 nm, the loss of $C_{gc}$ is ~8 % at the multilayer and ~2 % at 1L. For EOT = 1 nm, the loss of $C_{gc}$ becomes ~20 % at the multilayer and only 6 % at 1L. This trend has been experimentally confirmed by *C-V* measurements with more than 10 devices, as shown in **Figure 5c**. Although the variation is relatively large from sample to sample, the monolayer $MoS_2$ saturation capacitance is unambiguously larger than that of the thicker $MoS_2$. The above study indicates that monolayer 2D materials should be selected to maximize the carrier density at the accumulation region as well as the gate modulation ability since $n = \int C_{gc} dV_{TG}$ and $C_{gc}$ is largest for monolayer. This is the big advantage of the 2D layered channel compared with SOI because thickness scaling of SOI to the ultimate regime (~1 nm) negatively impact the electrical transport properties, i.e., the drastic degradation of the mobility [6,7]. Of course, monolayer 2D materials also suffer from the interfacial issues, such as the remote phonon scattering from the high-*k* oxide and Coulomb scattering due to charged impurities. These interfacial issues become more unavoidable because the carriers in the channel are close to the top gate oxide interface and even from the bottom oxide interface [32-34]. However, due to the inherent structural stability resulted from the layered structure provides the relatively high mobility even in the monolayer limit. By improving these interfacial issues, the ideal carrier electrostatics will be achieved by the gate electrical field.

## 3. Conclusion

In this study, the accumulation region of the *C-V* characteristics was systematically investigated for top-gate $MoS_2$ FETs with the thickness from 1L to 16 nm. A strong quantum-mechanical effect due to the channel thickness confinement was experimentally demonstrated in monolayer $MoS_2$ FETs, which was supported by the self-consistent calculation. At the thickness scaling limit to the monolayer $MoS_2$, $C_{gc}$ is maximized to be $C_{ox}$ because $C_A$ can be neglected due to the negligible channel thickness. Therefore, monolayer 2D channels are suggested to be used to enhance the maximum on-current as well as the gate modulation ability.

## Acknowledgements

N. F. was supported by a Grant-in-Aid for JSPS Research Fellows from the JSPS KAKENHI. This research was partly supported by The Canon Foundation, the JSPS Core-to-Core Program, A. Advanced Research Networks, the JSPS A3 Foresight Program, and JSPS KAKENHI Grant Numbers JP16H04343, JP19H00755, and 19K21956, Japan.